**Distinction of disorder, classical and quantum vibrational contributions to atomic mean-square amplitudes in dielectric pentachloronitrobenzene**

Jacqueline M. Cole,[*,1,2,3] Hans-Beat Bürgi,[4] Garry J. McIntyre[5,†]

[1]*Cavendish Laboratory, University of Cambridge, J. J. Thomson Avenue, Cambridge, CB3 0HE. UK*

*Departments of Chemistry[2] and Physics[3], University of New Brunswick, P. O. Box 4400, Fredericton, New Brunswick, E3B 5A3. Canada.*

[4]*Laboratorium für Kristallographie der Universität Bern, Freiestrasse 3, CH-3012 Bern, and Institute of Organic Chemistry, University of Zürich, Winterthurerstrasse 190, CH-8057 Zurich, Switzerland.*

[5]*Institut Laue-Langevin, B. P. 156, 38042 Grenoble Cedex 9, France.*

Corresponding Author:   Jacqueline M. Cole,
                        Cavendish Laboratory,
                        University of Cambridge,
                        J. J. Thomson Avenue,
                        Cambridge, CB3 0HE. UK.
                        Tel: +44 1223 337470
                        Fax: +44 1223 350266
                        Email: jmc61@cam.ac.uk

**Abstract**

The solid-state molecular disorder of pentachloronitrobenzene (PCNB) and its role in causing anomalous dielectric properties are investigated. Normal coordinate analysis (NCA) of atomic mean square displacement parameters (ADPs) is employed to distinguish disorder contributions from classical and quantum mechanical vibrational contributions. The analysis relies on multi-temperature (5-295K) single-crystal neutron diffraction data. Vibrational frequencies extracted from the temperature dependence of the ADPs are in good agreement with THz spectroscopic data. Aspects of the static disorder revealed by this work, primarily tilting and displacement of the molecules, are compared with corresponding results from previous, much more in-depth and time-consuming Monte Carlo (MC) simulations; their salient findings are reproduced by this work, demonstrating that the faster NCA approach provides reliable constraints for the interpretation of diffuse scattering. The dielectric properties of PCNB can thus be rationalised by an interpretation of the temperature dependent ADPs in terms of thermal motion and molecular disorder. The use of atomic displacement parameters in the NCA approach is nonetheless hostage to reliable neutron data. The success of this study demonstrates that state-of-the-art single-crystal Laue neutron diffraction affords sufficiently fast the accurate data for this type of study. In general terms the validation of this work opens up the field for numerous studies of solid-state molecular disorder in organic materials.

## I. INTRODUCTION

Molecular disorder in a crystal structure is often a crucial factor in lending molecular materials their desired physical property. As a consequence, there is a fundamental need to understand the detailed nature of molecular disorder. This paper presents a novel method to discriminate features of molecular orientational (or occupational) disorder from those of thermal motion using multiple-temperature, single-crystal neutron diffraction. The inclusion of data at 5K in this parametric study also enables the distinction of classical and quantum vibrational contributions to thermal motion.

Historically, rigid-body and segmented-rigid-body analysis of anisotropic displacement parameters (ADPs) or rigid-body refinements of crystal structures, at a single temperature, present a simple means by which one can partition ADPs of each atom into contributions from translation, libration and soft molecular deformation modes, at least approximately.[1-9] Rigid-body and semirigid-body models are relatively successful because they make reasonable assumptions about the correlations between displacements of different atoms. In general these assumptions are insufficient, however, to obtain the full matrix of mean square displacement amplitudes; a single-temperature crystal structure refinement has no recourse to some of the off-diagonal matrix elements, in particular those representing screw coupling of translation and libration or coupling between external and internal degrees of freedom.[10,11] Yet, such correlation parameters are necessary to define fully the relative phases of atomic displacements, together with a measure of the extent of atomic displacement that emanates from the diagonal matrix elements. Furthermore and most important in the present context, analysis of ADPs pertaining to a single temperature cannot distinguish between contributions from static disorder and thermal vibrations.[12]

### A. Multi-temperature Normal Coordinate Analysis

There is currently no general solution to the problems of distinguishing between static disorder and thermal motion and of determining the relative phases of atomic motions if only Bragg data at a single temperature are available. However, a Normal Coordinate Analysis (NCA) approach, developed by Bürgi and co-workers[13] can overcome both problems for simple molecules such as benzene, urea,[14] hydroquinone,[15] hexamethylenetetramine[16], naphthalene[17], and tris-ethylenediamine zinc sulfate [(H$_2$NCH$_2$CH$_2$NH$_2$)$_3$Zn] SO$_4$.[18] Key to this approach is the exploitation of a temperature parametric. ADPs from multiple-temperature crystallographic Bragg data sets provide a trend, $U_{ii}(T)$, that can be compared with that expected for a collection of simple harmonic oscillators. The provision of a wide temperature range of data is of course important to ensure a good comparison in the classical temperature regime; but equally crucial is the access to sufficiently low temperature data where quantum effects imply a deviation from classical behavior. It is the difference between quantum and classical behavior that enables one to extract the off-diagonal correlation elements of the mean-square displacement amplitudes. Such methods represent a very attractive alternative to a full lattice dynamics calculation (e.g. Refs 19,20) and may replace these numerically intense and time-consuming calculations to a certain extent.

Libration, translation and deformation modes occur in concert with each other and often have similar frequencies, but can be resolved on the basis of the temperature dependence of the ADPs. The NCA approach resolves such modes by exploiting the fact that the general formulation of the ADPs, $<u^2(i,T)> = \Sigma_j a_j(i)(\hbar/2\omega_j)\coth(\hbar\omega_j/2k_BT)$ for all normal modes j, with frequency $\omega_j$ and relative contribution $a_j(i)$ of i atoms, is a non-linear function of T. ADPs in the high ($T \to \infty$) and low ($T \to 0$) temperature limiting regimes are thus linearly independent. The off-diagonal elements of the mean-square displacement matrix, *i.e.* products $<u(i,T)_i u(i',T)_j>$ for different atoms i and i' moving in the same or in different directions, follow.

The considerations outlined so far apply to the vibrational modes that are of sufficiently low frequency to be significantly excited in the temperature range for which ADPs are available. In practice a temperature-independent contribution has to be considered for every atom. If this contribution is small compared to the atomic zero-point amplitudes from libration and translation, it can safely be assigned to the zero-point vibrational amplitudes of the high-frequency vibrations. If it is large, it is indicative of some kind of molecular disorder. Thus, molecular translation, libration, deformation and their correlations, as well as terms able to account for disorder, are incorporated into the NCA approach.

The analysis procedure employs a non-linear least-squares refinement of the multiple-temperature ADP data, where the weighted sum, $\Sigma_{i,j,k} w_{ij}[U_{ij}(i,T)_{obs} - U_{ij}(i,T)_{calc}]^2$ is minimised by allowing various parameters to vary from a starting model. The essential refinement variables are the effective normal mode frequencies $\omega_j$, the relative atomic displacements representing the respective eigenvectors (related to $a_j(i)$) and temperature-independent terms $\varepsilon_{ij}(i)$. Furthermore, where data deviate from harmonic behavior in the high end of the temperature range, a Grüneisen parameter, $\gamma^G$, can also be included to measure the level of anharmonicity associated with the thermal expansion of the material.[16]

In order that one can employ this temperature-parametric approach to the maximum possible molecular complexity within the confines of determinacy, the number of refinable parameters is minimized, *e.g.* by capitalizing on crystal symmetry wherever possible. In addition, one makes judicious assumptions about soft molecular deformations that are likely to contribute most significantly to atomic motion (usually of terminal atoms). The essence of the rigid-body and semirigid-body treatment of Schomaker and Trueblood[11] is not only embodied into this approach, it is also generalized. The new approach – essentially a normal coordinate analysis (NCA) based on the temperature evolution of the ADPs - partitions ADPs into the temperature-dependent components of translation, libration, internal motion, coupling between these, the small, nearly temperature-independent, components from high frequency vibrations and, most importantly in the present example, the temperature-independent contributions from disorder.

Given the greater sophistication and more fruitful prospects of this approach compared with previous rigid-body methods, one might question why its use has not proliferated more widely than it has currently. One of the key factors that has held up progress is the limited availability of neutron diffraction data over multiple temperatures, including at least one temperature point that pertains to the quantum regime.

## B. The key role of Laue-based neutron diffraction data

The NCA approach is predisposed to neutron diffraction data since the modeling of molecular disorder is very sensitive to atomic accuracy; in this regard, neutron data is superior to the much more commonly used X-ray diffraction analogue. This is because there is a systematic error associated with the assumption of the theoretical, free-atom form factor model used in the X-ray structure solution:

$$F(\boldsymbol{h}) = \sum_{i=1}^{N} f_i(\boldsymbol{h}) T_i(\boldsymbol{h}) e^{2\pi i \boldsymbol{h}.\boldsymbol{r}_i}$$

where $F(\boldsymbol{h})$ is the structure factor of reflection, $\boldsymbol{h}$ = (hkl); $f_i(\boldsymbol{h})$ is the atomic scattering amplitude for atom i, calculated from free-atom wavefunctions (the free-atom form factor); $T_i(\boldsymbol{h})$ is the Debye-Waller factor, and $\mathbf{r}_i$ is the mean position of atom, i.

The free-atom form factor assumes a spherical distribution of electrons. However, chemical bonding distorts this distribution, especially so in terminal atoms.[21] Consequently, the electronic charge density is often shifted away from the positions of the atomic nuclei. Since X-rays scatter off electrons rather than off the nucleus as in neutron diffraction, positional parameters of terminal atoms derived by X-ray diffraction reflect the charge-density centroid of these atoms rather than their true nuclear positions. One is thus very reliant on a theoretical model being appropriate to the problem in hand; otherwise, there may be a systematic error associated with $\mathbf{r}_i$. In stark contrast, $f_i(\boldsymbol{h})$ is a known constant for any given nucleus in neutron diffraction, thereby permitting the derivation of accurate atomic positional parameters $\mathbf{r}_i$ without recourse to a theoretical model for $f_i(\boldsymbol{h})$.

Due to the diffuse nature of electrons, the radial dependence of $f_i(\boldsymbol{h})$ is not the same for a given atom in different local environments and so $f_i(\boldsymbol{h})$ is not easily separable from $T_i(\boldsymbol{h})$ when interpreting X-ray structure factors. Consequently, it is not possible to evaluate $f_i(\boldsymbol{h})$ and $T_i(\boldsymbol{h})$ independently without the use of a theoretical model for $f_i(\boldsymbol{h})$. Any error in the radial characteristics of the free-atom form factor model for $f_i(\mathbf{h})$ propagates into the thermal motion parameters rendering them artificially larger than their true values when employing X-ray diffraction.[21,22] Once again, thermal parameters derived from neutron diffraction are unaffected and so should usually be significantly more accurate than their X-ray derived counterparts.[22]

While there are of course neutron facilities to which one can apply to obtain neutron diffraction data, the NCA approach requires several data sets measured over a range of temperatures. Multi-temperature crystal structure determinations can be rather time consuming, and the time efficiency of an experiment is a particularly pertinent consideration where the use of a scarce and expensive resource such as a neutron facility is involved. What is needed is a means by which one can obtain accurate multiple-temperature neutron data sets in rapid succession. Fortunately, the single-crystal Laue diffractometer, VIVALDI, at the ILL, Grenoble, France[24,25], is a perfect instrument for this purpose. Laue, rather than monochromatic diffraction, is an attractive prospect for rapid data collection because the diffraction condition is met many times simultaneously and so many reflections can be measured at the same time. However, the accuracy of results derived from Laue diffraction has, historically, suffered due to the incomplete coverage of reciprocal space arising from partial spatial overlapping of reflections and chromatic overlap, particularly at low sinθ/λ, and the added uncertainty of wavelength-dependent corrections. Thus, monochromatic experiments have hitherto generally prevailed as the choice of neutron diffraction technique where accuracy is a premium. That said, significant advances have been made recently to circumvent these problems and thus improve on the accuracy of the results (e.g. Refs. 26,27). The development of the VIVALDI instrument[24,25] is one such advance.

VIVALDI is sited on a thermal neutron guide, and the axis of the cylindrical detector is vertical which allows the mounting of a He cryostat to permit sample temperatures down to 1.5 K routinely. Since entering operation at the end of 2001, VIVALDI has already demonstrated its worth in rapid multiple-temperature crystallography: the variation of the Cr-O bond lengths in the ammonium chromium Tutton's salt $(ND_4)_2Cr(D_2O)_6(SO_4)_2$ was followed as a function of temperature in a series of six-hour data collections, each of four Laue patterns.[28] The observed variation of the bond lengths could be modelled by assuming a vibronic coupling model that had been obtained by fitting a Jahn-Teller Hamiltonian to the low-energy excitations of $[Cr(OH_2)_6]^{2+}$. The accuracy of ADPs is much more sensitive to experimental deficiencies than is the accuracy of bond geometry. However, a variety of statistical tests have shown that ADPs for all atoms can be well determined on an instrument like VIVALDI; see for example, the case study on zinc (tris)thiourea sulfate.[29]

### C. The case study of pentachloronitrobenzene

This paper seeks to bring together these innovations in molecular disorder modelling and neutron diffraction instrumentation, and to assess their combined potential, via the multiple-temperature neutron diffraction study of pentachloronitrobenzene (PCNB). This material is one in a family of chloronitrobenzenes that have a propensity to molecular disorder. PCNB exhibits solid-state molecular disorder and order-disorder phase transitions that have intrigued researchers for nearly half a century (*e.g*. refs 30-34). The fact that such disorder appears to be responsible for unusual dielectric behaviour[35-38] adds a substantial materials-centred interest to the problem. The results of this case study are shown to rationalize the role of molecular disorder on the dielectric properties observed.

From a technical point of view, PCNB represents the most far-reaching application of ADP-modeling to the separation of molecular motion from disorder. The average structure of the compound displays six-fold molecular orientational disorder that can be modeled adequately only if ADPs on each atom are well determined; such sensitivity makes this detailed analysis particularly rigorous. A diffuse scattering study,[39] that employs Monte Carlo modeling, has already been conducted on PCNB, and so their results act as an ideal comparative study to the current work.

## II. EXPERIMENTAL

A 2.5 x 1.0 x 1.0 mm$^3$ single crystal of PCNB, glued to a vanadium pin, was mounted inside a helium-flow cryostat on the Very-Intense Vertical-Axis Laue Diffractometer (VIVALDI) at the Institut Laue-Langevin (ILL), Grenoble, France. VIVALDI employs Laue diffraction using a polychromatic thermal-neutron beam and a large solid-angle (8 sterad), cylindrical image-plate detector,[24,25] to increase the detected diffracted intensity by one-to-two orders of magnitude, compared with a conventional monochromatic experiment. Four Laue diffraction patterns, each separated by 30° intervals in rotation of the crystal perpendicular to the incident neutron beam, were collected with 2.5-3 hour acquisition time per pattern, over a 90° range, at three temperatures: T=100.0(4)K, 200.0(7)K and 295.0(10)K. At T = 5.00(2)K, the same procedure was followed, except that data were collected over a 180° range, to give seven Laue diffraction patterns in total. The enhanced data acquisition at 5K is important for this study because this temperature point lies at one extreme of the temperature range studied and since a temperature-parametric approach is adopted in the modelling, this temperature point will influence the nature of the least-squares fit more than other values used in the fitting process – mathematically, such points are said to have a high leverage.

The Laue patterns were indexed using the program LAUEGEN of the Daresbury Laboratory Laue Suite,[40,41] and the reflections integrated using the local program INTEGRATE+ which uses a two-dimensional version of the three-dimensional minimum σ(I)/I algorithm.[42] The individual reflections were corrected for absorption using the calculated (wavelength-dependent) absorption coefficient, $0.02747\lambda + 0.007773$ mm$^{-1}$ (transmission range: 0.885 – 0.962). The reflections were normalised for the incident wavelength, using an empirical curve derived by comparing equivalent reflections and multiple observations, via the program LAUENORM.[43] All reflections in the wavelength range 1.1Å to 2.5Å were accepted in this procedure and thence used in the structural refinements. The coverage at all temperatures was around 76% of the unique reflections to the minimum *d* spacing observed (0.68 Å at 5 K; 0.74 Å at 100, 200 and 295 K). This level of coverage has been demonstrated to provide reliable anisotropic displacement parameters on VIVALDI, in comparison to monochromatic data.[29] The data coverage yields an observation-to-parameter ratio of 7.6-9.8 which is reasonably well distributed in direction within the unique volume of reciprocal space after folding in the trigonal symmetry.

The model was adjusted by full-matrix least-squares refinement using SHELXL-97.[44] Since the Laue technique only allows ratios of the linear lattice parameters to be determined, initial cell parameters from X-ray diffraction were assumed or extrapolated in the refinement of the neutron data. Rigid-bond tests were undertaken using PLATON.[45] Positional and anisotropic displacement parameters were refined for all atoms, except the nitrogen atom, where positional and isotropic displacement parameters only were refined. This atom-selective difference in treatment of displacement parameters was applied in order to control the disorder model within sensible means, since the disordered nitrogen and chlorine atoms lie so close to each other. A full summary of crystal, data collection and refinement details, atomic fractional coordinates and bond distances and angles is archived in the supplementary information[46].

## III. RESULTS AND DISCUSSION

### A. Conventional structural analysis

The average structure of PCNB displays $R\bar{3}$ space-group symmetry with parallel layers of apparently coplanar molecules stacked along the threefold axes of symmetry (Fig. 1). Each molecule exhibits six-fold orientational disorder where the nitro group is equally likely to lie at any one of the six aromatic substitution sites. This is deduced from the disorder model that is the best least-squares fit to the Bragg data and features 1/6:5/6 $NO_2$:Cl occupation factors at each substitution site. Fig. 2 shows ADP ellipsoid plots of the structure for each temperature in mean molecular plane projection. Fig. 3 shows that the oxygen atoms of the nitro group are rotated out of the plane of the rest of the molecule, to the extent that the two mean planes are nearly perpendicular to each other (74(1)° apart). The ADPs on the oxygen atoms reveal a distinct torsional or 'wagging' motion, or possibly a combination of the two, while the displacement ellipsoids of carbon and chlorine are noticeably elongated in the out-of-plane direction, i.e. the $U_{33}$ values are much larger than $U_{11}$ and $U_{22}$ at all temperatures. Table I lists the ADP magnitudes, $U_{ii}$, with 2σ uncertainty, for all atoms.

An assessment of the rates of change of $U_{ii}$ values with temperature (Fig. 4) can provide some insight into the apparent libration effects of the atoms in PCNB. Such rates for each atom will be near identical if only translational motion exists, whilst atoms undergoing mostly librational motion will display a markedly higher rate if they are further from the libration axes (Cl vs C). Fig. 4 (insert) reveals that small libration effects are observable in the $U_{11}$ and $U_{22}$ directions, respecting the relative statistical uncertainties; the chloro group appears to librate significantly in the $U_{33}$ direction which lies perpendicular to the molecular plane.

The onset of non-classical behaviour is very apparent for all atoms in the 5K data by virtue of the deviation from linearity at low temperature with an intercept above that expected from linear extrapolation of the high temperature data to 0 K. The $U_{ii}$ value at the y-axis intercept provides a rough estimate of temperature-independent effects for each atom. Zero-point motion is usually the dominant contribution to this quantity. However, the case of PCNB is rather peculiar since the temperature-independent quantities for certain atoms (in particular directions), especially O1 and O2 ($U_{33}$) and Cl ($U_{33}$), are far too large to represent molecular zero-point mean square displacements.[15]

A more quantitative assessment of the nature of the various contributions to the mean square amplitudes is possible via Normal Coordinate Analysis.[13] This approach provides a fuller characterisation of the molecular disorder in PCNB, including the separation between disorder and thermal motion, the implicit modelling of interatomic (but not intermolecular) correlations and the explicit partitioning of librational and translational vibrations of the molecules. The simultaneous least-squares refinement of data from all temperatures in the NCA approach is statistically superior to conventional structural analysis where, in this case study, the data interpretation at a single temperature is statistically limited.

### B. Normal coordinate analysis and comparison with spectroscopic data

The ADPs of the disordered molecules were interpreted in two steps within the NCA approach described earlier.[13] In a first approximation, $C_6[Cl_{5/6}(NO_2)_{1/6}]_6$ was considered as if it were $C_6Cl_6$. Due to populations of 1 and 5/6, ADPs of C and Cl are well determined experimentally and the molecular symmetry of $C_6Cl_6$ is close to 6/mmm. This implies that a small number of variable parameters should be sufficient to model the ADPs satisfactorily. The model established for temperature-dependent and independent contributions comprised two librations and one translation out of the plane ($l_x$, $l_y$, $t_z$), a libration and two translations in the plane ($l_z$, $t_x$, $t_y$), two temperature-independent tensors: $\varepsilon_{ij}(C)$ and $\varepsilon_{ij}(Cl)$, and a Grüneisen parameter to account for anharmonicity due to thermal expansion. The agreement between the model with 17 parameters and the 48 experimental observations (12 $U_{ij}$s at each temperature) is satisfactory with $R2 = [\Sigma(U_{obs}-U_{calc})^2/\Sigma(U_{obs})^2]^{1/2} = 0.029$. The quantity $[\Sigma(U_{obs}-U_{calc})^2/(n_{obs}-n_{par})]^{1/2} = 0.0015$ Å$^2$ is in reasonable agreement with standard uncertainties of the $U_{ij}$ (see Table I).

In a second step of the interpretation, a $C_6Cl_5NO_2$ molecule was considered. One of the Cl atoms and its ADPs were replaced by a $NO_2$ group. The model for temperature dependent and independent contributions remained unchanged except for two key additions: (i) three additional, temperature independent terms are included: matrices $\varepsilon_{ij}(O1)$, $\varepsilon_{ij}(O2)$ and $\varepsilon_{iso}(N)$ with local x-axes from O1 to O2, local y-axes from C to N and local z-axes perpendicular and right-handed for O1, left-handed for O2; (ii) a torsional vibration about the C-N bond was introduced in order to account for extra motion of the $NO_2$ group; it was allowed to couple with overall translations and librations. It transpired that the only significant coupling was found between this torsion and the libration $l_z$. The fit between experiment and model is somewhat worse than for the first model, $R2 = 0.063$. However, the contributions to the quantity $[\Sigma(U_{obs}-U_{calc})^2]$ are very uneven, 0.0004 Å$^2$ for $C_6Cl_5$, practically the same as for the $C_6Cl_6$ model, but 0.0021 Å$^2$ for the $NO_2$ group. The standard uncertainties of the $U_{ij}$s of the $NO_2$ group are also larger, however, (see Table I) due to the much smaller populations (0.167, 1 and 0.833 for O, C and Cl, respectively). Thus the increase in R2 is due entirely to the lower quality of the ADPs of the $NO_2$ group and the use of the same unit weight for all observations. Note that unit weights were used in the least-squares calculations to account for the unknown reliability of our model which was developed for ordered structures, but is applied here to a disordered structure.

The salient results from the two ADP analyses are broadly similar (see Table II). A comparison between the temperature-dependent parts of the analyses shows that the libration and translation frequencies differ only marginally, except for $l_z$. These marginal differences are due primarily to the changes in the inertial moments on going from $C_6Cl_6$ to $C_6Cl_5NO_2$ and affect the calculated ADPs insignificantly. The decrease of the libration frequency $\omega(l_z)$ by ~10 cm$^{-1}$ arises from coupling with the NO$_2$ torsional mode.

The librational frequencies derived from the ADPs (43.3(1.8), 48.3(2.3), 34.4(1.5)) provide an excellent match to THz spectroscopy data whose lowest frequency band is broad and centred at 40 cm$^{-1}$; it has been attributed to whole-molecule libration.[47] The torsional frequency derived from our model is 69(3) cm$^{-1}$. Further work by Reid and Evans[48] reveals spectral peaks in the region 75-110 cm$^{-1}$, which are partly attributed to torsional NO$_2$ motion. The peaks in this solid-state spectrum are red-shifted in the analogous solution spectrum; this suggests that torsional NO$_2$ motion is affected by intermolecular interactions.

The average translational frequencies of $C_6Cl_5NO_2$ may be compared to those of $C_6D_6$ (ref. 14) which crystallizes in an ordered structure:

$$f(C_6Cl_5NO_2) / f(C_6D_6) = [\nu(C_6Cl_5NO_2) / \nu(C_6D_6)]^2 [M(C_6Cl_5NO_2)/M(C_6D_6)]$$

where the $f$'s are translational force constants and the M's molecular masses. With the average translation frequencies of 32 and 46 cm$^{-1}$, respectively, the ratio becomes 1.4, implying the reasonable result that the intermolecular forces are stronger in the chloro than in the hydrocarbon compound. Anharmonicity due to thermal expansion of the PCNB crystal led to a Grüneisen parameter of reasonable magnitude 2.9(3) [*c.f.* 3.5(2) for hexamethylenetetramine[16]].

## C. Disorder analysis and comparison with diffuse scattering studies

As for the temperature independent parts of the two ADP analyses, the results from both models are the same within experimental error (see Table II). The temperature-independent contributions to the ADPs of C and Cl are represented by (nearly diagonal) tensors $\varepsilon_{ij}$ with respective mean square displacements of 0.0068(3) Å$^2$ and 0.0095(3) Å$^2$ along the C-Cl bonds, and 0.0069(3) Å$^2$ and 0.0068(4) Å$^2$ perpendicular to the C-Cl bonds within the molecular plane. The displacements in z are much larger, 0.0367(4) Å$^2$ for C and 0.0712(4) Å$^2$ for Cl. Thus the temperature-independent terms $\varepsilon_{33}$ contribute more than 90% to the large observed values of $U_{33}$, at least at 5 K. They are far too large to represent molecular zero-point mean square displacements arising from molecular deformation vibrations.[15] They are therefore interpreted as being due primarily to positional and tilting disorder of the molecules. The two components may be separated by considering $\varepsilon_{33}$(C) and $\varepsilon_{33}$(Cl) as sums of a mean square translational displacement $\varepsilon_{33}$(trans) along z and mean square out-of-plane tilt displacements $\varepsilon_{33}$(tilt), which are proportional to the square of the distance, d, of C and Cl, respectively, from the centre of the molecule:

$$\varepsilon_{33} = \varepsilon_{33}(\text{trans}) + \varepsilon_{33}(\text{tilt}) = <\Delta_{33}^2> + <\delta\alpha^2> d^2.$$

with d(C) = 1.4 and d(Cl) = 3.1 Å, $<\Delta_{33}^2>$ is 0.0258 Å$^2$ and $<\delta\alpha^2>$ is 0.00545 rad$^2$ (corresponding to 4.2°).

The relative signs of the molecular shift and tilt cannot be obtained from mean-square values alone. They do follow, however, from the displacement of the N atom or the midpoint between the two oxygen positions from the mean $C_6Cl_6$ plane. As there is only a single $NO_2$ group in the molecule, the six-fold orientational disorder does not superimpose several such groups and therefore leads to real rather than averaged atomic positions. With $d(N) = 2.86$ Å, $-<\Delta_{33}^2>^{1/2} = -<0.0258 \text{ Å}^2>^{1/2} = -0.161$ Å and $<\delta\alpha^2>^{1/2} = <0.00545 \text{ rad}^2>^{1/2} = 0.0738$ rad, the N-atom would be displaced 0.05 Å from the molecular plane, in agreement with the observed value of 0.06(1) Å. Correspondingly, estimated and observed values for the O…O midpoint [d(midpoint O…O) = 3.25 Å] are 0.08 and 0.11(1) Å, respectively. The agreement is judged sufficient to justify the above choice of sign and the average magnitudes of the tilt and displacement of PCNB from its average position.

Structurally the signs of $<\Delta_{33}^2>^{1/2}$ and $<\delta\alpha^2>^{1/2}$ may be interpreted as follows. A PCNB molecule is related to its neighbours across centres of inversion (at least in the average structure). Its $NO_2$ group points to, and is nearly coplanar with a C-Cl bond of a molecule in the next layer up of the $R\bar{3}$ packing (Fig. 1). It is tilted away from it by 4.2°, but moves its centre of mass towards it by 0.16 Å (probability near 1; the presence of a second C-$NO_2$ group across the inversion centre would imply an unreasonably short O…O non-bonded distance of ~2.6 Å). Correspondingly, the C-Cl bond *para* to the $NO_2$ group points at a C-Cl bond of a molecule in the next layer down of the $R\bar{3}$ packing (probability 0.833) or a C-$NO_2$ group (probability 0.167). The juxtaposition of a C-$NO_2$ group and a Cl-C bond would imply an O…C non-bonded contact of >2.8 Å if tilt and displacement are born in mind, somewhat short, but not unreasonably so for an oxygen atom with partial negative charge in the neighbourhood of a carbon atom with partial positive charge as it bears an electronegative substituent. The C…Cl contact is unproblematic at >3.8 Å.

With regard to the tensors $\varepsilon_{ij}(O1)$ and $\varepsilon_{ij}(O2)$, these both show very large components in the direction perpendicular to the $NO_2$ plane, ~0.07 Å$^2$, corresponding to static displacements of the O1 and O2 atoms from the $NO_2$ plane by about 0.26 Å, and quite comparable to the displacements of the C, N and Cl atoms from the mean molecular plane arising from the overall molecular tilt and displacement as described above. To keep the rigid $NO_2$ group planar the oxygen atoms must displace to opposite sides of this plane indicating a static torsional disorder of the $NO_2$ group by ~14° in addition to the dynamic torsional vibration.

While corresponding parameters associated with the N atom were also calculated in this study, they are not interpreted. This is because only isotropic displacement parameters were available for the N atom from the structure refinement of neutron diffraction data; the close proximity of the N atom with occupation 1/6 to the Cl with occupation 5/6 occludes the ADPs of the nitrogen atom[49], thereby restricting its parameterisation in the structure refinement. Nonetheless, the inclusion of the N atom in the calculations was important in order to frame the nitro group in the model.

The interpretation of the temperature independent ε-tensors may be compared with results from a corresponding diffuse scattering study on PCNB that employed MC modelling.[39] Despite extensive efforts to simulate the disorder as occupancy short-range order using MC methods, these all failed which led to the conclusion that the nitro group exhibits six-fold orientational disorder around the benzene ring with a random distribution. This finding, together with the observation of large ADPs associated with $NO_2$ and Cl, yielded the hypothesis that the diffuse scattering emanates entirely from atomic displacements, induced to cope with the large differences in size between $NO_2$ and Cl. The supramolecular stability of a molecule is particularly vulnerable to such size disparities since the latter can result in uncomfortably close intermolecular interactions if there is no means of compensation for size effects. It transpired that MC simulations with very large displacements readily explained both strong and broad diffuse-scattering signatures, thus corroborating the hypothesis. The atomic positions from the final MC model revealed that the $NO_2$...$NO_2$ groups tilt away from each other while Cl...Cl atoms bend towards each other. Associated intermolecular separations deviated from those of the average structure by +0.6 Å (N...N), +0.3 Å (N...Cl) and -0.08 Å (Cl...Cl).

These salient conclusions are the same as those obtained in this study, thereby validating the NCA analysis. The tilt-displacement model implies shifts of the Cl atoms from their mean position in the average structure by between -0.06 to 0.48 Å; for comparison the displacements found in the much more involved study of diffuse scattering are between -0.08 and 0.59 Å.[39] The agreement between the two results suggests that the disorder parameters found from the analysis of the temperature evolution of the ADPs may serve as a guide or even constraint in the initial stages of interpreting diffuse scattering data which is known to be computationally expensive.

**D Relationships between the molecular disorder and the dielectric properties of PCNB**

Solid-state molecular rotation has long been hypothesized to be responsible for the anomalous dielectric properties of PCNB. Aihara and co-workers[50] evidenced this influence via their dielectric absorption measurements as a function of frequency (30 Hz – 1 MHz) and temperature (293-372 K). Distinct anomalous dispersion of the dielectric absorption was observed and was shown to shift towards a higher frequency with increasing temperature. A temperature-dependent form of molecular rotation within the molecule was consequently suspected. They proposed that the associated rotation axis was perpendicular to the molecular plane (in-plane rotation). On the basis of dipole measurements, they deduced that the rotational energy barriers were less than 180° apart. Hall and Horsefall[32] confirmed these findings and also showed that the supramolecular environment can alter substantially the dielectric properties of PCNB. This was revealed by a comparative dielectric absorption study of solid-state PCNB in the crystalline form versus that of PCNB incorporated in a non-polar matrix of naphthalene. When PCNB lay in the host matrix, dielectric absorption profiles were shifted markedly to higher frequencies at a given temperature; this implies that the molecule becomes freer to rotate with the removal of the crystal lattice. In addition, the distribution parameter decreased from 0.3 to 0.05, *i.e.* towards a single relaxation time; a promotion of free-rotation would manifest this change. Together, these dielectric observations indicated that some form of rotational motion may exist in the crystalline lattice but it is hindered somewhat by intermolecular interactions.

Aihara and co-workers[50] also calculated the energies associated with various rotational origins of this suspected motion in order to try to identify the true nature of rotational motion associated with the observed dielectric behaviour. The energy barrier calculated for full rotation of the molecule about the axis perpendicular to the ring was $\Delta H = 5634$ cm$^{-1}$; this corresponds well with more recent Monte Carlo (MC) cluster calculations: $\Delta H = 5601$ cm$^{-1}$.[51] An activation energy of $\Delta H = 2759$ cm$^{-1}$ pertaining to the full rotation of the NO$_2$ group was also calculated by these MC calculations. These values were suspiciously high as they afforded unreasonable entropy changes and issues with the Onsager equation.[50]

Our NCA results corroborate the Monte Carlo simulations of diffuse scattering[39] in showing that the NO$_2$ does not undergo any form of atomic site rotation. Rather, the NO$_2$ groups are substitutionally disordered, manifesting large static atomic displacements as well as dynamic torsional NO$_2$ vibrations. The static displacements are spatially constrained by intermolecular interactions. This model explains the observed dielectric properties in terms of its marked dependence on both supramolecular environment and temperature. In this latter respect, given that the dielectric properties were measured in the temperature range, T = 293-372 K, anharmonic effects may also play a role, given the NCA derivation of the Grüneisen parameter, $\gamma^G = 2.9(3)$.

## IV. CONCLUDING REMARKS

Our results have identified and modelled structural disorder in an organic material; it is due to highly-correlated intermolecular NO$_2$...Cl interactions which prevent an entirely ordered PCNB crystal lattice. The interactions are evidenced by anomalously large, temperature-independent, disorder-related atomic displacements amounting to more than 90% of the atomic displacement parameters (ADPs) at 5 K. The Normal Coordinate Analysis (NCA) employed to identify these displacements also separates them from classical and quantum mechanical vibrational displacements that are due to whole-molecule libration and translation and internal nitro-group torsions.

The NCA analysis is very attractive on account of the modest investment in data analysis compared with the well-established alternatives that are very time-consuming and impose heavy demands on calculations: primarily this refers to Monte Carlo (MC) modelling of diffuse scattering as well as full lattice dynamics calculations for a heavily disordered crystal. Since attractiveness is nonetheless hostage to the assurance of good accuracy in the results, at the current stage of development, it is important to verify the results of the NCA analysis. MC modelling provides the natural verification procedure since the accepted model is chosen on the basis of its level of ability to simulate the diffuse-scattering signatures in the experimental data; pending a good match, one is therefore confident of the MC-generated structural model.

Regarding an overall comparison, it is worth observing that the multi-temperature aspect of NCA analysis affords information about molecular disorder that is inaccessible to the MC modelling of diffuse scattering data pertaining to a single temperature. In particular, NCA analysis allows the explicit separation of libration and translation vibrational contributions per atom and associated enumeration of their modal frequencies; the distinction between temperature-independent and temperature-dependent contributions to the atomic displacements; and the quantification of anharmonic effects. The two approaches are therefore highly complementary.

For future work on PCNB, the mode frequencies determined from ADPs provide important preliminary information for inelastic neutron scattering experiments that will permit the evaluation of force and elastic constants in PCNB, and the dispersive influence of its $NO_2$...Cl intermolecular interactions. In turn, this stands to yield further valuable insight into the role of structure and dynamics in the observed dielectric properties in PCNB.

From an experimental basis, the success of the NCA analysis also acknowledges that the use of VIVALDI to generate the multi-temperature neutron diffraction data required for this type of analysis is entirely adequate. Therefore, modern Laue neutron diffraction provides a fast and accurate means by which NCA analyses on further compounds can be realised. It is noteworthy to add that radial projection in the Laue technique leads to partial overlap of diffuse and Bragg features, and possibly an underestimation of the $U_{ij}$s. However since no distinct diffuse scattering features were observed elsewhere on these Laue patterns we can assume that there was no artifactual increase in the intensity of the Laue spots. Improvements to the upstream guide and replacement of the image plates since this experiment has led to a 10-fold improvement in detection efficiency on VIVALDI.[25] This would enable the same experiment to be performed faster or on a smaller sample or with more intermediate temperature points

The normal coordinate interpretation of the diffraction data has provided a new insight into the origins of the anomalous dielectric absorption in PCNB. The ability to enumerate frequencies of individual vibrational modes has been particularly helpful in this regard, since one can compare the results with previous spectroscopic findings. While there are previously reported examples of the application of this NCA approach, this is the first to relate its results to a real-world physical property; in the process this has answered a long-standing question about the origin of the anomalous dielectric characteristics of PCNB. This work therefore opens up serious prospects for solving other materials-science problems. In the present study, this approach has disentangled structural disorder, due to highly correlated intermolecular atomic interactions, from molecular motion. There is already a range of indirect evidence for such correlations being responsible for a wide range of interesting electrical, optical and magnetic physical properties in organic materials, *e.g.* in organic superconductivity,[52] luminescence,[53] non-linear optics,[54] photovoltaics,[55] and ferromagnetism[56]. The further application and development of the NCA approach is therefore poised to further a deeper understanding of such physical phenomena.

## ACKNOWLEDGEMENTS

The authors wish to acknowledge the Institut Laue-Langevin, Grenoble, France, for access to central facilities beamtime. JMC is also indebted to the Royal Society for a University Research Fellowship, the University of New Brunswick, Canada, for The UNB Vice-Chancellor's Research Chair, and the University of Bern for hosting her visit that allowed the NCA work to be undertaken.

## REFERENCES

\*   Author for correspondence. jmc61@cam.ac.uk

†   Present address: Bragg Institute, Australian Nuclear Science and Technology Organisation, Locked Bag 2001, Kirrawee DC NSW 2232, Australia.

1   D. W. J. Cruickshank (1956). Acta Cryst. **9**, 757-758.

2   D. W. J. Cruickshank (1956). Acta Cryst. **9**, 1005-1009.

3   V. Schomaker, and K. N. Trueblood (1968). Acta Cryst. B**24**, 63-76.

4   C. K. Johnson (1970). Thermal Neutron Diffraction, edited by B. T. M. Willis. Oxford University Press.

5   J. D. Dunitz, and D. N. J. White (1973). Acta Cryst. A**29**, 93-94.

6   K. N. Trueblood (1978). Acta Cryst. A**34**, 950-954.

7   K. N. Trueblood, and J. D. Dunitz (1983). Acta Cryst. B**39**, 120-133.

8   X. –M. He, and B. M. Craven (1985). Acta Cryst. A**41**, 244-251.

9    X. –M. He, and B. M. Craven (1993). Acta Cryst. A**49**, 10-22.

10   H. B. Bürgi, (1989). Acta Cryst. B**45**, 383-390.

11   V. Schomaker, and K. N. Trueblood (1998). Acta Cryst. B**54**, 507-514.

12   K. N. Trueblood, H.B. Bürgi, H Burzlaff, J. D. Dunitz, C. M. Gramaccoli, H. H. Schulz, U. Shmueli, S. C. Abrahams. (1996) Acta Cryst. A52, 770-781.-

13   H.B. Bürgi, and S.C. Capelli (2000). Acta Cryst. A**56**, 403-412.

14   S. C. Capelli, M. Förtsch, and H.B. Bürgi (2000). Acta Cryst. A**56**, 413-424.

15   H.-B. Bürgi, and S.C. Capelli (2003). Helv. Chim. Acta **86**, 1625-1640.

16   H.B. Bürgi, S.C. Capelli, and H. Birkedal (2000). Acta Cryst. A**56**, 425-435.

17   S. C. Capelli, A. Albinati, S. A. Mason, and B. T. M. Willis (2006). J. Phys. Chem. A **110**, 11695-11703.

18   S. Smeets, P. Parois, H.-B. Bürgi and M. Lutz (2011). Acta Cryst. B**67**, 53–62.

19   M. Born, and K. Huang, (1954). Dynamical Theory of Crystal Lattices. Oxford: Clarendon Press.

20   A. A. Maradudin, E. W. Montroll, G. H. Weiss, and I. P. Ipatova (1971). Theory of Lattice Dynamics in the Harmonic Approximation, 2nd ed. New York: Academic Press.

21   P. Coppens, and C. A. Coulson, (1967). Acta Cryst. **23**, 718-720.

22   W. C. Hamilton (1969). Acta Cryst. A**25**, 194-206.

23   P. Coppens, (1969). Acta Cryst. A**25**, 180-186.

24   C. Wilkinson, J. A. Cowan, D. A. A. Myles, F. Cipriani, and G. J. McIntyre (2002). Neutron News **13**, 37-41.

25   G. J. McIntyre, M.-H. Lemeé-Cailleau, and C. Wilkinson (2006). Physica B **385-386,** 1055-1058.

26   D. A. Keen, M. J. Gutmann, and C. C. Wilson (2006). J. Appl. Cryst. **39**, 714-722.

27   Z. Ren, D. Bourgeois, J. R. Helliwell, K. Moffat, V. Srajer, and B. L. Stoddard (1997). J. Synchrotron Rad. **6**, 897-917.

28  C. Dobe, C. Noble, G. Carver, P. L. W. Tregenna-Piggott, G. J. McIntyre, A.-L. Barra, A. Neels, S. Janssen, and F. Juranyi (2004). J. Am. Chem. Soc. **126**, 16639-16652.

29  J. M. Cole, G. J. McIntyre, M. S. Lehmann, D. A. A. Myles, C. Wilkinson, and J. A. K. Howard, (2001). Acta Cryst. A**57**, 429-434.

30  C. A. Meriles, S. C. Perez, and A. H. Brunetti (1996). Phys. Rev. B **54** 7090-7093.

31  N. T. Correia, J. J. M. Ramos and H. P. Diogo (2002). J. Physics and Chemistry of Solids, **63**, 1717-1722.

32  P. G. Hall, and G. S. Horsefall (1973). J. Chem. Soc. Faraday II **69**, 1071-1077.

33  I. Tanaka, F. Iwasaki, and A. Aihara (1974). Acta Cryst. B**30**, 1546-1549.

34  T. Sakurai (1962). Acta Cryst. **15**, 1164-1173.

35  H. A. Kolodziej, K. Orzechowski, R. Szostak, P. Freundlich, T. Glowiak, S. Sorriso (1996). J. Mol. Struct. **380**, 15-22.

36  H. A. Kolodziej, P. Freundlich, S. Sorriso, C. Sorriso (1999). Chem. Phys. Lett. **305**, 375-380.

37  C. Kitazawa, and I. G. Tandai (1979). Abs. Papers. Am. Chem. Soc., 172.

38  K. Saito, H. Kobayashi, Y. Miyazaki, and M. Sorai (2001). Solid State Comm. **118**, 611-614.

39  L. H. Thomas, T. R. Welberry, D. J. Goosens, A. P. Heerdegan, M. J. Gutmann, S. J. Teat, P. L. Lee, C. C. Wilson, J. M. Cole (2007). Acta Cryst. B**63**, 663-673.

40  J. W. Campbell, (1995). J. Appl. Cryst. **28**, 228-236.

41  J. W. Campbell, Q. Hao, M. M. Harding, N. D. Nguti, and C. Wilkinson (1998). J. Appl. Cryst. **31,** 23-31.

42  C. Wilkinson, H. W. Khamis, R. F. D. Stansfield., and G. J. McIntyre (1988). J. Appl. Cryst. **21,** 471-478.

43  J. W. Campbell, J. Habash, J. R. Helliwell, and K. Moffat (1986). Information Quarterly for Protein. Crystallogr. **18**, 23-31.

44  G. M. Sheldrick (1997). SHELXL-97. A program for the refinement of crystal structures. University of Göttingen, Göttingen, Germany.

45  A. K. Spek, (2001). "PLATON, a multipurpose crystallographic tool", Utrecht University, Utrecht, The Netherlands. http://www.cryst.chem.uu.nl/platon/.

46  See supplementary material at EPAPS WEBSITE for full crystallographic information files of the neutron diffraction refinements of PCNB at each temperature.

47  C. Reid, G. J. Evans, M. W. Evans (1979). Spectrochimica Acta **35**A 679-680.

48  C. J. Reid, M. W. Evans (1980). Molecular Physics **41** 987-998.

49  While the neutron scattering lengths of N (9.36 barns) and Cl (9.58 barns) are comparable, the ratio of 5:1 Cl: N renders the Cl as the dominant scatterer.

50  A. Aihara, C. Kitazawa, A. Nohara, (1970). Bull. Chem. Soc. Japan **43,** 3750-3754.

51  L. H. Thomas (2007). PhD Thesis. University of Cambridge, UK.

52  J. M. Williams, M. A. Beno, H. H. Wang, P. C. W. Leung, T. J. Emge, U. Geiser and K. D. Carlson, Acc. Chem. Res. 18 (1985) 261-267.

53  J. Cornil, A. J. Heeger, J. L. Bredas, Chem. Phys. Lett. 272 (1997) 463-470.

54  N. R. Behrnd, G. Labat, P. Venugopalan, J. Hulliger and H. B. Bürgi, *Cryst. Eng. Comm.* 12 (2010) 4101-4108.

55  P. E. Keivanidis, I. A. Howard, R. H. Friend, Adv. Funct. Mater. 2008, 18, 3189–3202

56  D. Belo, M. J. Figueira, J. P. M. Nunes, I. C. Santos, L. C. Pereira, V. Gama, M. Almeida and C. Rovira, J. Mat. Chem. 16 (2006) 2746-2756.

**Table I**. Equivalent ($U_{eq}$) and principal components ($U_{ii}$) of the anisotropic displacement parameters (Å² x 10³) for pentachloronitrobenzene at T = 5K, 100K, 200K and 295K, for all atoms except nitrogen; isotropic displacement parameters, $U_{iso}$, for nitrogen. Uncertainties are 2σ($U_{ii}$)

|  | $U_{ii}$ | C | Cl | O1 | O2 | N |
|---|---|---|---|---|---|---|
| **5.00(2) K** | $U_{11}$ | 9(1) | 10(1) | 18(4) | 10(3) | - |
|  | $U_{22}$ | 9(1) | 9(1) | 56(7) | 78(9) | - |
|  | $U_{33}$ | 39(1) | 78(2) | 52(4) | 34(5) | - |
|  | $U_{eq}$ | 19(1) | 33(1) | 42(3) | 42(3) | - |
|  | $U_{iso}$ | - | - | - | - | 20(2) |
| **100.0(4) K** | $U_{11}$ | 14(1) | 16(1) | 36(6) | 14(5) | - |
|  | $U_{22}$ | 16(1) | 19(1) | 87(11) | 104(16) | - |
|  | $U_{33}$ | 48(1) | 89(3) | 52(7) | 57(8) | - |
|  | $U_{eq}$ | 26(1) | 41(1) | 55(4) | 67(6) | - |
|  | $U_{iso}$ | - | - | - | - | 34(4) |
| **200.0(7) K** | $U_{11}$ | 23(1) | 26(1) | 30(6) | 22(5) | - |
|  | $U_{22}$ | 25(1) | 35(1) | 113(13) | 141(16) | - |
|  | $U_{33}$ | 60(1) | 114(3) | 75(7) | 82(10) | - |
|  | $U_{eq}$ | 36(1) | 59(1) | 76(5) | 88(5) | - |
|  | $U_{iso}$ | - | - | - | - | 52(5) |
| **295.0(10) K** | $U_{11}$ | 34(1) | 39(1) | 55(9) | 44(7) | - |
|  | $U_{22}$ | 39(1) | 55(1) | 151(17) | 210(20) | - |
|  | $U_{33}$ | 79(1) | 151(4) | 89(9) | 98(10) | - |
|  | $U_{eq}$ | 51(1) | 82(2) | 101(6) | 121(8) | - |
|  | $U_{iso}$ | - | - | - | - | 68(7) |

**Table II.** Librational, translational and torsional motions in PCNB as deduced from normal coordinate analysis of $U_{ij}$ data from 5, 100, 200 and 295K data sets.

| | $l_x$ | $l_y$ | $l_z$ | $t_x$ | $t_y$ | $t_z$ | torsion |
|---|---|---|---|---|---|---|---|
| $C_6Cl_6$ | (cm$^{-1}$) | (cm$^{-1}$) | (cm$^{-1}$) | (cm$^{-1}$) | (cm$^{-1}$) | (cm$^{-1}$) | (cm$^{-1}$) |
| | 43.8(7) | 43.8(7) | 44.3(1.0) | 35.2(4) | 35.2(4) | 32.4(5) | - |
| | | | | | | | |
| | $\varepsilon(C)$ | (10$^4$ Å$^2$) | | | $\varepsilon(Cl)$ | (10$^4$ Å$^2$) | |
| | 68(3) | 0(2) | 0(2) | | 95(3) | 1(2) | 0(2) |
| | | 69(3) | 0(2) | | | 68(2) | 0(2) |
| | | | 367(4) | | | | 712(4) |
| | | | | | | | |
| $C_6Cl_5NO_2$ | (cm$^{-1}$) | (cm$^{-1}$) | (cm$^{-1}$) | (cm$^{-1}$) | (cm$^{-1}$) | (cm$^{-1}$) | (cm$^{-1}$) |
| | 43.3(1.8) | 48.3(2.3) | 34.4(1.5) | 33.3(9) | 33.3(9) | 29.5(1.0) | 69.2(2.6) |
| | | | | | | | |
| | $\varepsilon(C)$ | (10$^{-4}$ Å$^2$) | | | $\varepsilon(Cl)$ | (10$^{-4}$ Å$^2$) | |
| | 63(8) | 0(6) | 3(6) | | 91(9) | 1(5) | -28(6) |
| | | 63(8) | 0(6) | | | 44(12) | -7(6) |
| | | | 358(10) | | | | 724(11) |
| | | | | | | | |
| | $\varepsilon(O1)$ | (10$^{-4}$ Å$^2$) | | | $\varepsilon(O2)$ | (10$^{-4}$ Å$^2$) | |
| | 212(18) | 89(15) | 198(14) | | 405(18) | -53(15) | 204(14) |
| | | 177(15) | 128(14) | | | 39(15) | 1(14) |
| | | | 683(26) | | | | 705(27) |
| | | | | | | | |
| | $\varepsilon_{iso}(N)$ | (10$^4$ Å$^2$) | 216(15) | | | | |

**Figure captions**

**FIG. 1:** Partial packing diagrams of PCNB. This illustrates the (Top) near-parallel arrangement of the C-(Cl)$_{5/6}$(NO$_2$)$_{1/6}$ residues between neighbouring layers in the average structure (arrangement of the NO$_2$-groups arbitrary), and (Bottom) the slight tilt of the (blue) nitrogen atom from the mean molecular plane of the lower molecule away from the one above.

**FIG. 2.** The disordered arrangement of the NO$_2$ group over six positions in the crystal structure of pentachloronitrobenzene. Atoms are represented with 50% probability ADP ellipsoids at 5, 100, 200 and 295K, projected onto the mean molecular plane.

**FIG. 3:** An out-of-plane projection of pentachloronitrobenzene at 5K.

**FIG. 4.** Variation in the anisotropic displacement parameter magnitudes, $U_{ii}$, in pentachloronitrobenzene as a function of temperature. It is noteworthy to add that the $dU_{ii}/dT$ values were calculated from the gradient of the least-squares linear fit to the 100K, 200K and 295K data. This is because one can reasonably assume that, to a first approximation, an atom exhibits the classical behaviour of a simple harmonic oscillator in this temperature region, which follows a linear trajectory. All $dU_{ii}/dT$ values, except for $U_{11}$ O(1), $U_{11}$ O(2) and $U_{22}$ O(2), are statistically significant according to a t-test with 90% confidence interval; standard deviations of the gradient are annotated to the Figure insert to further illustrate the good reliability of most rates. Overall, the linear model represents the data well, as evidenced by the coefficients of determination ($R^2$) that range from 0.92-0.99 for all linear least-squares fits except for $U_{11}$ O(1) where $R^2 = 0.52$. Three of the twelve y-axis intercepts differ significantly from zero according to a t-test with 90% confidence interval: $U_{33}$[Cl, 0 K] = 0.055(9) Å$^2$, $U_{33}$[O(1), 0K] = 0.034(5) Å$^2$, and $U_{33}$[O(2), 0K] = 0.037(5) Å$^2$.

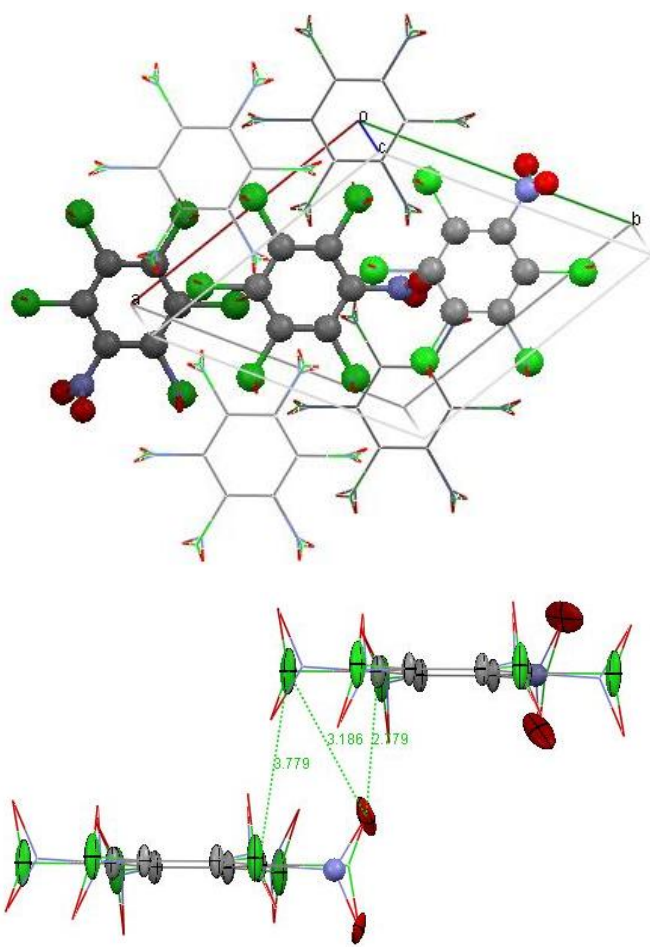

**FIG. 1.** Partial packing diagrams of PCNB. This illustrates the (Top) near-parallel arrangement of the C-(Cl)$_{5/6}$(NO$_2$)$_{1/6}$ residues between neighbouring layers in the average structure (arrangement of the NO$_2$-groups arbitrary); (Bottom) slight tilt of the (blue) nitrogen atom from the mean molecular plane of the lower molecule away from the one above.

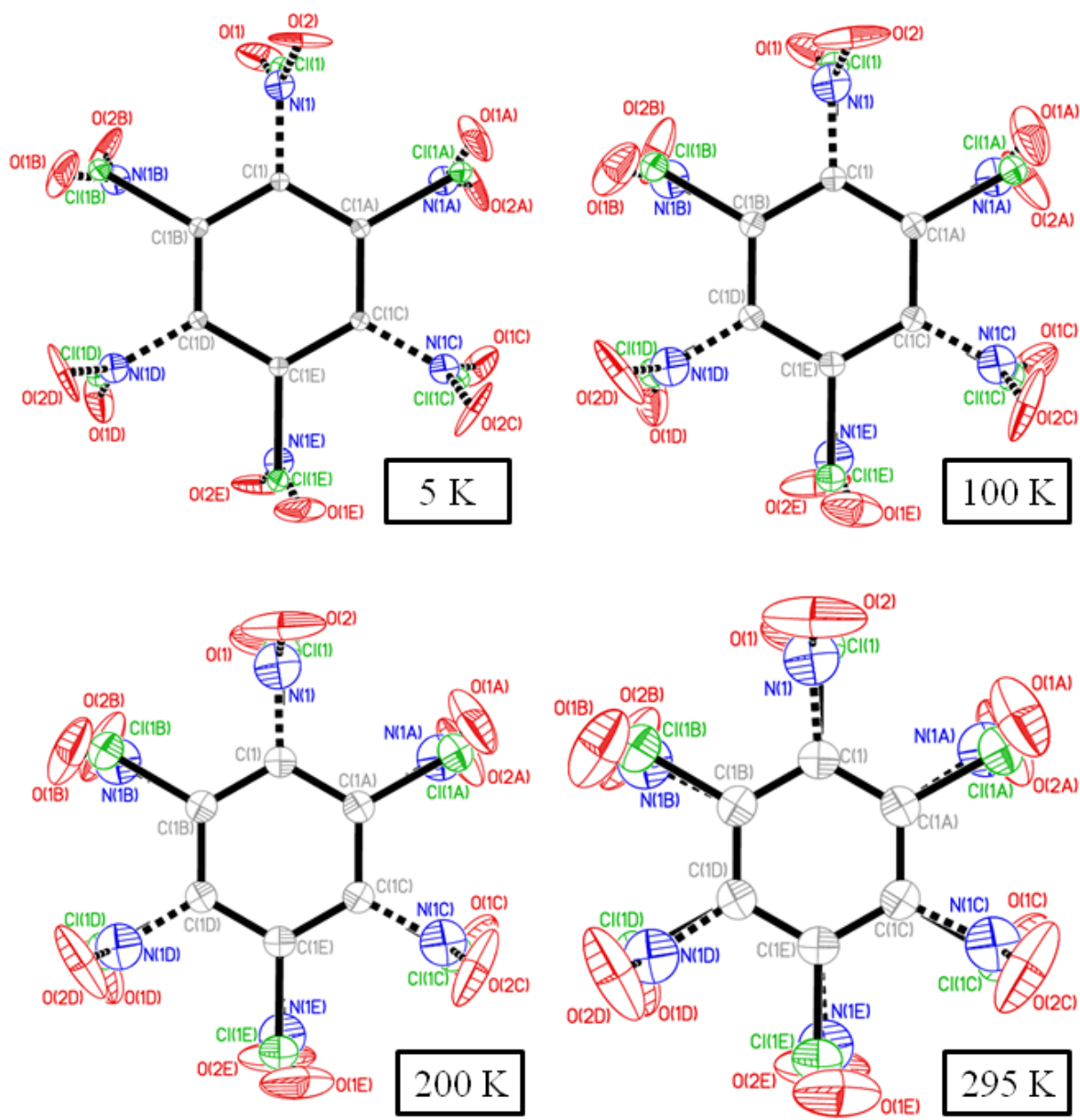

**FIG. 2.** The disordered arrangement of the NO$_2$ group over six positions in the crystal structure of pentachloronitrobenzene. Atoms are represented with 50% probability ADP ellipsoids at 5, 100, 200 and 295K, projected onto the mean molecular plane.

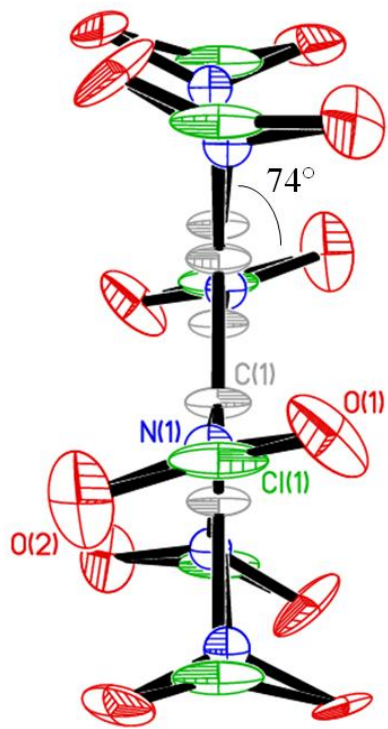

**FIG. 3.** An out-of-plane projection of pentachloronitrobenzene at 5K.

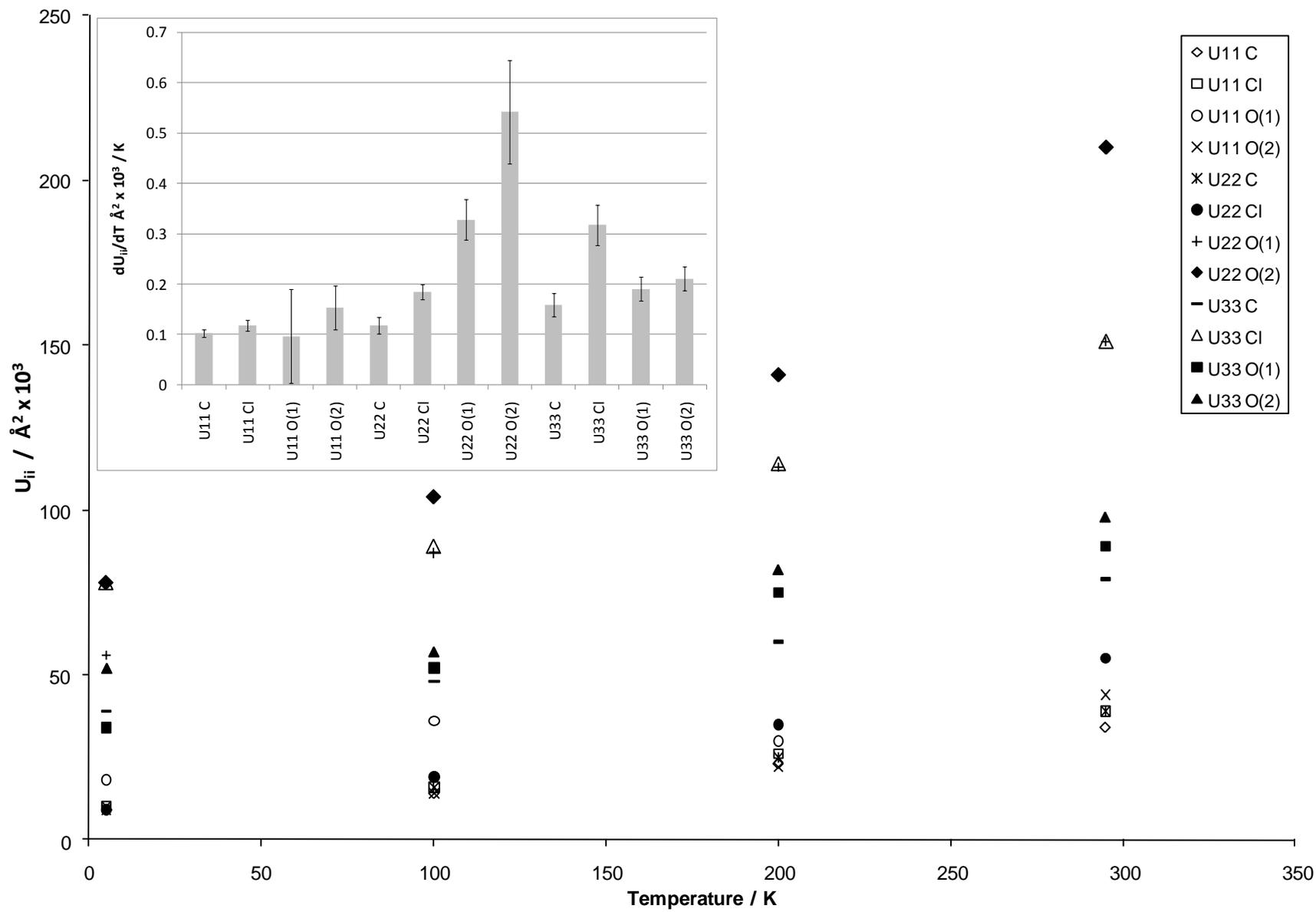

**FIG. 4.** Variation in the anisotropic displacement parameter magnitudes, $U_{ii}$, in pentachloronitrobenzene as a function of temperature. It is noteworthy to add that the $dU_{ii}/dT$ values were calculated from the gradient of the least-squares linear fit to the 100K, 200K and 295K data. This is because one can reasonably assume that, to a first approximation, an atom exhibits the classical behaviour of a simple harmonic oscillator in this temperature region, which follows a linear trajectory. All $dU_{ii}/dT$ values, except for $U_{11}$ O(1), $U_{11}$ O(2) and $U_{22}$ O(2), are statistically significant according to a t-test with 90% confidence interval; standard deviations of the gradient are annotated to the Figure insert to further illustrate the good reliability of most rates. Overall, the linear model represents the data well, as evidenced by the coefficients of determination ($R^2$) that range from 0.92-0.99 for all linear least-squares fits except for $U_{11}$ O(1) where $R^2 = 0.52$. Three of the twelve y-axis intercepts differ significantly from zero according to a t-test with 90% confidence interval: $U_{33}$[Cl, 0 K] = 0.055(9) Å$^2$, $U_{33}$[O(1), 0K] = 0.034(5) Å$^2$, and $U_{33}$[O(2), 0K] = 0.037(5) Å$^2$.